\documentclass[10pt]{article}
\usepackage{multirow}
\usepackage[dvips]{graphicx}

\usepackage[psamsfonts]{amssymb}
\usepackage{amsxtra}
\usepackage{threeparttable}
\usepackage{amsmath}
\usepackage{amssymb}

\usepackage{graphicx}

\usepackage{cite}

\usepackage{color}

\usepackage{setspace}
\usepackage{float}

\usepackage{url}

\usepackage[labelfont=bf,labelsep=period,justification=raggedright]{caption}

\makeatletter
\renewcommand{\@biblabel}[1]{\quad#1.}
\makeatother

\date{May 15, 2013}

\begin{document}

\begin{flushleft}
{\Large
\textbf{Novel Virtual Moving Sound-based Spatial Auditory Brain--Computer Interface Paradigm}\footnote{The final publication is available at IEEE Xplore \url{http://ieeexplore.ieee.org} and the copyright of the final version has been transferred to IEEE \copyright2013}

}
Yohann Leli\`evre$^{1,2,3}$,
and Tomasz M. Rutkowski$^{2,3,}$\footnote{The corresponding author. E-mail: \url{tomek@tara.tsukuba.ac.jp}}
\\
\bf{1} Department of Neurosciences and Neuropsychopharmacology, University of Bordeaux, France\\
\bf{2} Life Science Center of TARA, University of Tsukuba, Tsukuba, Japan\\
\bf{3} RIKEN Brain Science Institute, Wako-shi, Japan\\
E-mail: \url{tomek@tara.tsukuba.ac.jp}\\
\url{http://about.bci-lab.info/}
\end{flushleft}

\section*{Abstract}

This paper reports on a study in which a novel virtual moving sound-based spatial auditory brain--computer interface (BCI) paradigm is developed. Classic auditory BCIs rely on spatially static stimuli, which are often boring and difficult to perceive when subjects have non--uniform spatial hearing perception characteristics. The concept of moving sound proposed and tested in the paper allows for the creation of a P300 oddball paradigm of necessary target and non--target auditory stimuli, which are more interesting and easier to distinguish. We present a report of our study of seven healthy subjects, which proves the concept of moving sound stimuli usability for a novel BCI. We compare online BCI classification results in static and moving sound paradigms yielding similar accuracy results. The subject preference reports suggest that the proposed moving sound protocol is more comfortable and easier to discriminate with the online BCI.

\noindent{\bf Keywords:} auditory BCI, P300, EEG, neurotechnology

\section{Introduction}

Severe motor disabilities can limit a person's ability to communicate, especially in patients suffering from amyotrophic--lateral--sclerosis (ALS), severe cerebral palsy, head trauma, multiple sclerosis, or muscular dystrophies. Such patients are incapable of conveying their intentions (locked--in syndrome) to the external environment. Considering that ALS is the third most common neurodegenerative disease, with an incidence of $5$ in $100,000$ cases per year in Japan, and that this disease mostly occurs in adulthood~\cite{alsKihira2005}, the improvement in patients' dependence on health care support is a major issue for aging societies. Over the last decades, numerous research projects have been undertaken in order to develop novel communication techniques which could rehabilitate or bypass the peripheral nerves and muscles destroyed by disease degenerative processes. One promising method utilizes central neural system activities, recorded by electroencephalography (EEG), to establish a new communication channel, as in a brain--computer interface application~\cite{bciWolpaw2000}. 
An exogenous BCI requires the user's sensory ability to be involved in a stimulating environment to induce sensory neurophysiological responses. 

So far, the primary choice of interaction modality has been vision. Most of the current BCI systems rely on the ability of the subject to control eye movements. However, the patients' inability to gaze directly, adjust focus, or blink their eyes may make the use of the visual modality in the BCI application very difficult. Therefore, other modalities are now being explored such as hearing~\cite{bciHill2005} and touch~\cite{bciMuller-Putz2006} in order to create vision independent BCIs.

The first auditory paradigm BCIs developed were based on binary decisions, allowing for lower information transfer rates (ITR) compared with multi--class interface solutions. In order to address the need for a multi--class BCI, a spatial auditory paradigm has been proposed~\cite{iwpash2009tomek,bciSchreuder2010,yohannMASTER2013}. This paradigm uses spatially distributed, auditory cues. Subjects just need to focus on the selected sound directions.  However this paradigm still does not produce ITR results leading to a fast BCI application that could rely solely on the auditory modality. Indeed its ITR is still inferior compared with the visual paradigms. 
Moreover, the problem of lower audible angles in spatial auditory perception limits the accuracy of static sound-based BCI applications. Thus, we propose the use of moving sound stimuli to create richer and more easily perceived auditory events~\cite{bciMoonJeong2013}. 

Our hypothesis can be summarized as follows. In spatial auditory BCI experimentation, the use of moving sound stimuli will solve the problem of lower audible angles and should lead to an increase in accuracy compared with classic static sound stimuli. In order to test this hypothesis, a simple five spatial vowel speller task is proposed. We test and compare the concept with seven healthy subjects by comparing their static and moving sound BCI spelling results.

The rest of the paper is organized as follows. In the next section, the experimental methods are described. This is followed by EEG signal analysis and classification of the results. The paper ends with conclusions and an outline of the direction of future research.

\section{Methods}

The experiments involved seven healthy subjects (mean age of $30.71$ years, with a standard deviation of $7.70$ years). All the experiments were performed at the Life Science Center of TARA, University of Tsukuba, Japan. The online EEG BCI experiments were conducted in accordance with \emph{The World Medical Association Declaration of Helsinki - Ethical Principles for Medical Research Involving Human Subjects}. The two protocols were designed to reproduce the auditory experiments as proposed in~\cite{bciMoonJeong2013}, with the addition of a moving sound modality.

The $200$~ms long spatial unimodal (auditory) stimuli were presented from five distinct virtual spatial locations through the use of headphones. All the subjects underwent a psychophysical test with a button press response to confirm understanding of the experimental set-up of the two protocols. These tests resulted in average response delays with means of around $450$~ms, meaning that the tasks were well understood and the mental load differences among the stimuli were not significant. The stimuli sounds used were the Japanese vowels \emph{a, i, u, e, o}, represented in \emph{hirigana}. During the online BCI experiments, the EEG signals were captured with eight active dry electrodes g.SAHARA, connected to a g.MOBIlab+ EEG amplifier by g.tec Medical Engineering GmbH, Austria. The electrodes were attached to the following head locations $Cz, CPz, P1, P2, P3, P4, Cp5,$ and $Cp6$, as in the $10/10$ extended international system~\cite{Jurcak20071600}. 
The ground and reference electrodes were attached to the left and right mastoids respectively. The recorded EEG signals were processed by the in-house enhanced BCI2000 application using a Stepwise Linear Discriminant Analysis (SWLDA) classifier with features drawn from $0$ to $700$~ms event-related potential (ERP) intervals. The sampling frequency was set to $256$~Hz, with the high pass filter at $0.1$~Hz, the low pass filter at $40$~Hz, and with a power line interference notch filter set in the $48-52$~Hz band. The Inter-Stimuli Interval (ISI) was set to $500$~ms and each stimuli length was $200$~ms. The subjects were instructed to spell the five vowel random sequences, which were presented audibly in each session. The target was presented with $20\%$ probability. It has been shown that this is rare enough to produce a clear P300 response~\cite{bciSellers2006}. Each target was presented ten times in a single spelling trial and the averages of ten ERPs were later used for the classification in order to make the experiment easier for novices. Both the static and moving sound stimuli oddball auditory trials were done three times by each subject.

\begin{figure}[b]
  \centering
  \includegraphics[height=10cm]{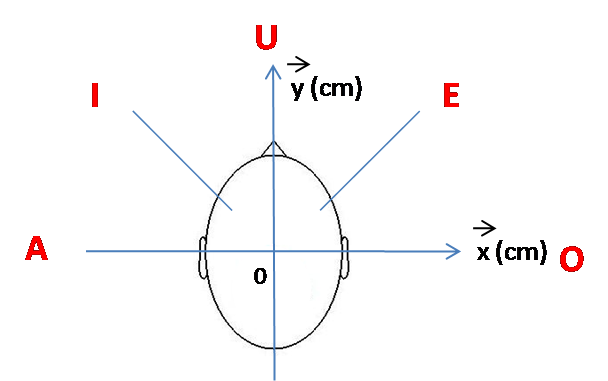}
  \caption{Auditory source directions visualization in the static sound BCI protocol.}\label{fig:StaticProtocol}
\end{figure}

In the static sound, the auditory source image of the vowel was virtually positioned at a distance of one meter from the subject using a vector based amplitude panning approach~\cite{nozomuANDtomekAPSIPA2012}. The five vowels were angularly separated by $45^\circ$ from right to left, as depicted in Figure~\ref{fig:StaticProtocol}.

\begin{figure}
  \centering
  \includegraphics[width=\linewidth]{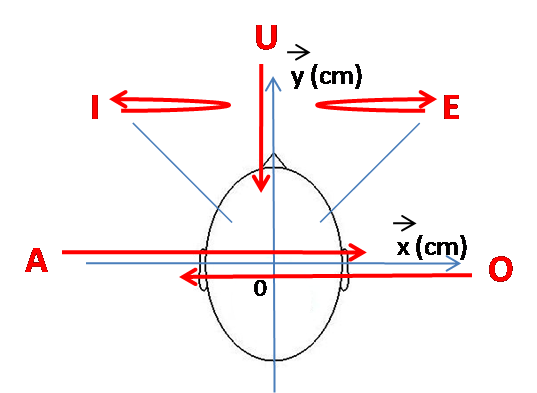}
  \caption{Auditory source directions and movement trajectories visualization in the moving sound BCI protocol.}\label{fig:MovingProtocol}
\end{figure}

In the moving sound protocol, the angle origins were the same as in the static example. The moving sound effects were applied in order to simulate five movement trajectories, as depicted in Figure~\ref{fig:MovingProtocol}.


\section{EEG Response Analysis and BCI Classification}

The segmented EEG data were first detrended and responses with an amplitude larger than 80~$\mu$V were eliminated from further analysis. The analysis of EEG data was based on online results obtained with a P300 Classifier of the BCI2000 environment. Also a subjective comfort preference questionnaire about static and moving sounds was completed by the subjects.

\begin{figure}[b]
  \centering
  \includegraphics[width=\linewidth]{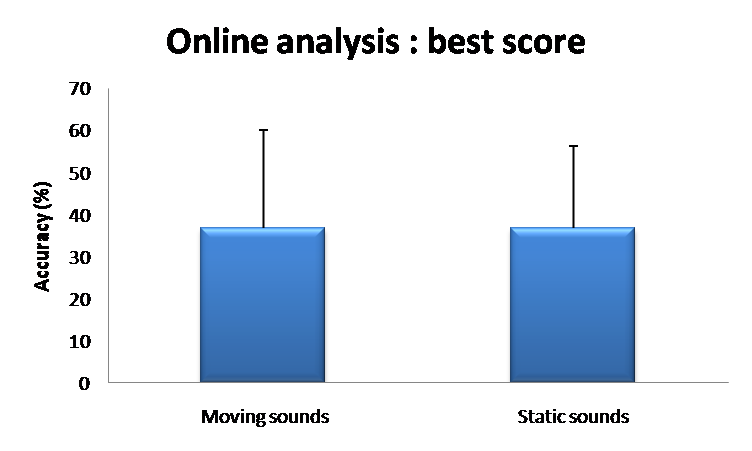}
  \caption{Online BCI classification results analysis comparison for the static and moving sound protocols.}\label{fig:OnlineAnalysis}
\end{figure}

The BCI classification results we obtained from online experiments, averaged for all subjects, are depicted in the Figure~\ref{fig:OnlineAnalysis}. In the pilot study, the same accuracy was obtained for static and moving sound stimuli. Yet the subjects reported the oddball P300 paradigm to be easier to identity with the moving sound stimuli.

\begin{figure}
  \centering
  \includegraphics[width=0.9\linewidth]{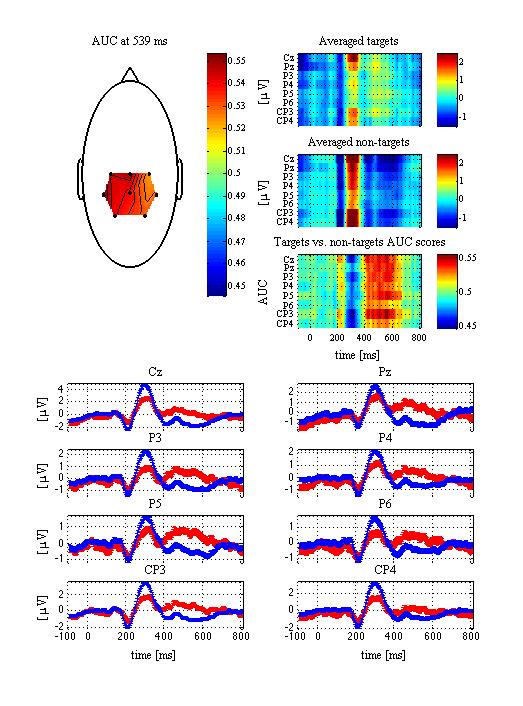}
  \caption{Auditory ERP visualization together with P300 responses obtained in the static sound spatial auditory BCI experiment.}\label{fig:AllStatic001}
\end{figure}

\begin{figure}
  \centering
  \includegraphics[width=0.9\linewidth]{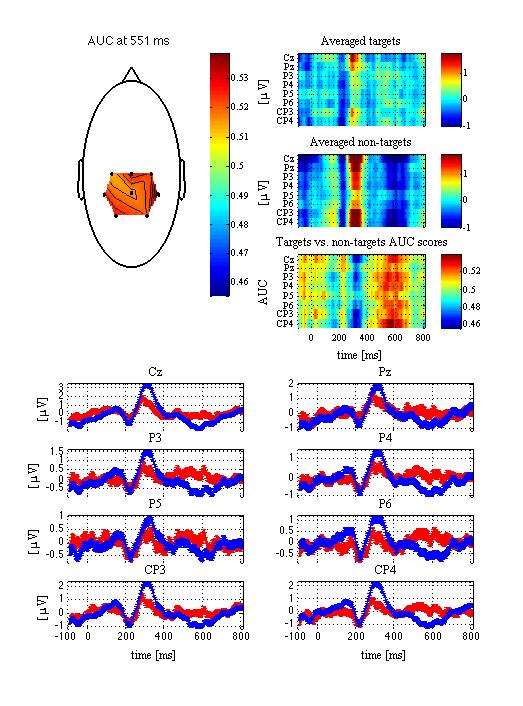}
  \caption{Auditory ERP visualization together with P300 responses obtained in the moving sound spatial auditory BCI experiment.}\label{fig:AllMove001}
\end{figure}

The P300 grand mean average results of response analysis for all subjects are depicted in Figures~\ref{fig:AllStatic001}~and~\ref{fig:AllMove001} for static and moving sound protocols respectively. The top--left panels in both figures represent the area under the curve (AUC) values projected on a head topographic plot with the EEG electrode locations used. The AUC method allows for ERP response discrimination analysis for subsequent classification, as also depicted in the three top--right panels in both figures for target (nota bene, P300 responses), non--target and AUC time series.

The best mean discrimination between target and non-target stimuli was obtained at $539$~ms after stimulus emission for the static sounds, and at $551$~ms after stimulus emission for the moving sounds. These periods fit in well with the rise in P300 in the case of auditory stimuli~\cite{bciHalder2013}. The AUC scores for static sound were between $0.52$ and $0.55$, which allows for the subsequent classification of features. For the moving sound, the scores similarly resulted in between $0.51$ and $0.54$.
The three top--right panels in Figures~\ref{fig:AllStatic001}~and~\ref{fig:AllMove001} represent the grand mean average responses of all subjects for target and non--target stimuli, and the AUC scores over time for each electrode. For the static protocol, we can see that the differences in electrical activity between target and non-target were the most significant for the $P200$~\cite{bciGolob2011}, around $320$~ms, and for the $P300$, around $540$~ms. Yet the AUC value is significant (above $0.5$) only for the $P300$ event. Similar responses were obtained for the moving sound protocol, as depicted in Figure~\ref{fig:AllMove001}, which is proof of the proposed concept.

The eight graphs on the bottom part of the two figures depict in details the evoked target (red) and non--target (blue) responses with clear P300 discrimination for each electrode separately.

\begin{figure}
  \centering
 \includegraphics[width=0.7\linewidth]{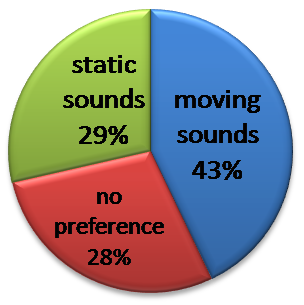}
  \caption{Stimulus preference and comfort analysis results for all subjects between static and moving protocols.}\label{fig:ComfortAnalysis}
\end{figure}

Even though the accuracy between static and moving sound protocols resulted in non--significant differences, we were able to observe the subjects' stimuli comfort preferences. Figure~\ref{fig:ComfortAnalysis} reports the subjects' preferences, which are greater for the moving ($43\%$) compared with the static ($29\%$), and no preference ($28\%$) results. The preference analysis also supports the proposed concept of the validity of a moving sound stimulus for spatial auditory BCI utilization.

\section{Discussion}

The aim of this experiment was to evaluate moving sound sources in comparison with classic protocols, in order to incorporate the concept in the spatial auditory BCI paradigm. Online results obtained with the SWLDA classifier show that there is no significant difference in accuracy. We also discovered that the $P200$ event might be used for classification. 

Even though there was no significant increase in accuracy of the moving sound in comparison to the static, the moving sound protocol still avoided the problem of too low audible angles in auditory spatial cognition. These are the novel and attractive points of the proposed new paradigm.

The moving sounds protocol was revealed to be more comfortable than the static one.  This characteristic is of interest because auditory paradigms tend to be more boring than other modalities.

This work will help to get closer to our objective of designing a more user friendly BCI, even if it does not fully attain this goal. Hence, we can expect that totally locked-in syndrome (TLS) patients will be able to use an appropriate BCI interface more efficiently and comfortably to restore their basic communication abilities.

\section{Conclusions}

Even if the results of this study do not lead directly to a new spatial auditory BCI paradigm and the restoration of the ability of TLS patients to communicate, they do demonstrate that the use of a moving sound stimulus improves the interface comfort, on the basis of subject preference reports. 
Indeed, even if the accuracy does not increase, our research proves that a moving sound stimulus is more comfortable than a static one. 
The proposed moving sound protocol helps to avoid the problem of lower audible angles in spatial auditory perception. Moreover, this work reveals another track to investigate, which might increase the ITR based on the $P200$ ERP latency. 

There still remains a long way to go before providing an efficient and comfortable auditory BCI, but our research has progressed toward this goal.

\section*{Acknowledgements}

We acknowledge the essential support of the International FidEx program of the University of Bordeaux France, the Student Exchange Support Program of the Japan Student Services Organization (JASSO), the Strategic Information and Communications R\&D Promotion Program (SCOPE) of the Ministry of Internal Affairs and Communications (MIC) of Japan, and Dr. Andrzej Cichocki from the RIKEN Brain Science Institute, Japan.
We also thank the YAMAHA Sound \& IT Development Division for the valuable technical support, and the members of the Multimedia Laboratory of the Life Science Center of TARA at University of Tsukuba, Japan, for their kind cooperation.



\end{document}